\documentclass[epj,nopacs]{svjour}
\usepackage{latexsym}
\usepackage{graphicx,amssymb,amsmath}
\usepackage[english]{babel}
\usepackage{appendix}
\usepackage{hyperref}\hypersetup{pdfstartview=FitH,pdfpagemode=None}
\usepackage{fancyhdr}
\begin{document}

\title{Observers in an accelerated universe.}

\author{Prado Mart\'{\i}n-Moruno}%
%
\institute{Colina de los
Chopos, Instituto de F\'{\i}sica Fundamental, \\
Consejo Superior de Investigaciones Cient\'{\i}ficas, Serrano 121,
28006 Madrid, Spain$^0$}
\date{Received: date / Revised version: date}
%
\abstract{If the current acceleration of our Universe is due to a cosmological constant,
then a Coleman-De Luccia bubble will nucleate in our Universe.
In this work, we consider that our observations could be likely in this framework, consisting
in two infinite spaces, if a foliation by constant mean curvature hypersurfaces is taken to count the events in the spacetime.
Thus, we obtain and study a particular foliation, which covers the existence of most
observers in our part of spacetime.
} 

\maketitle

\markboth{Prado Mart\'{\i}n-Moruno}{Observers in an accelerated universe}

\footnotetext{Present address: School of Mathematics, Statistics, and Operation Research,
 Victoria University of Wellington, PO Box 600, Wellington 6140, New Zealand.}

Our Universe is currently undergoing a period of accelerated
expansion. If this acceleration is due to a cosmological constant,
then our future Universe will be a de Sitter space. Coleman and De
Luccia considered that a de Sitter universe could be understood as
a false vacuum which would decay into the true vacuum
\cite{Coleman:1980aw} (see also Ref.~\cite{Coleman:1977py}),
leading to the nucleation of a bubble in the original de Sitter
space which grows at a velocity close to the speed of light
\cite{Coleman:1980aw}. Despite implying a catastrophic end for
part of the original universe, that process was gladly received in
the inflationary paradigm, residing at the very basis of Linde's
eternal inflation \cite{Linde:1986fc}, and it could also lead to
interesting future scenarios \cite{Garriga:1997ef}. Moreover,
those Coleman-De Luccia bubbles (CDL) have taken a renewed
interest (see, for example, Ref.~\cite{Yamauchi:2011qq}) in the
context of the string theory landscape \cite{Susskind:2003kw}. 

On the other hand, the nucleation of a CDL bubble is not the most 
surprising phenomenon which could take place in our future Universe.
As a de Sitter spacetime last forever, an infinite number of putative observers
could form from thermal and/or vacuum fluctuations arbitrarily late (see \cite{Page:2008zh} references therein);
therefore, it could seem that our observations are unlikely.
It should be worth noticed that such paradoxical result depends on the assigned measure and, at the end of the day, 
on how we count the events within the spacetime.
As it is well known, general relativity cannot provide us with a preferred
foliation by spacelike hypersurface in order to count the number of events within a spacetime. 
A possible choice is a
foliation by hypersurfaces of constant mean curvature (CMC), 
which can be considered in different spacetimes regardless of the particular symmetry. Furthermore,
those foliations provide us with a quantity which can play the
role of time \cite{York,Kuchar} and they can be
used to study different topics from canonical general relativity
\cite{Barbour:2010xk} to numerical relativity
\cite{Metzger:2004pr}.
Moreover, Page has pointed out that the foliations by CMC
hypersurfaces could also play an essential role regarding a new
approach to the measure problem in the framework of eternal inflation
\cite{Page:2008zh}, where the mentioned paradoxical result is also present. 
It has been suggested that in the case of a de Sitter space with a CDL bubble of de Sitter space inside 
this foliation might not be enough to cover the existence of most observers in this space. Nevertheless, 
if that would not be the case and a
foliation by CMC hypersurfaces could cover the existence of most
observers in our part of spacetime, then that could provide us
with at least approximately the right measure for our observations
\cite{Page:2008zh}.

It is the main aim of the present paper
considering whether a foliation by CMC hypersurfaces of a de Sitter space with a
CDL bubble of de Sitter space inside
could cover the existence of most observers in our
region of spacetime. It should be emphasized that this spacetime consists of two infinite spacetimes. Therefore,
if a CMC foliation covers most of the spacetime, then it can be used, by taking a certain measure, to get at least approximately 
the right probability to the occurrence of the events. 
That could also be done even in the case that the foliation may not penetrate significantly in the inside space,
if one assumes that the main contribution to obtain the probabilities would come from our spacetime region \cite{Page:2008zh}.
Nevertheless, if the foliation would not be enough to cover our existence,
then choosing such a foliation would imply that our observations are extremely unlikely, being more
natural putative observers measures. As that would be an uncomfortable result, in that case we should discard the use of CMC foliations.

\begin{figure}
\begin{center}
\includegraphics[width=0.8\columnwidth]{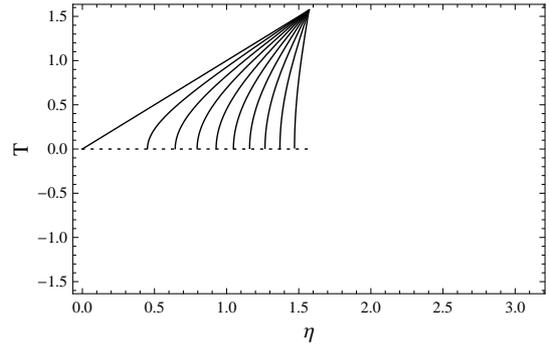}
\end{center}
\caption{Trajectory of the wall for different decay models corresponding to values of
$D/\alpha$ in the interval $[1,0.1]$, from the left to the right. The C-P diagram of the outside space is the
same as that of the de Sitter universe removing the region on the
left of the bubble wall from the moment that it nucleates, $T=0$. A similar figure can be obtained considering the inside space, corresponding the diagram to the region
bounded by $0<T_b<\pi/2$ and $0\leq\eta_b\leq \eta_b^*$.}
\label{diagtray}
\end{figure}

First of all, the spacetime must be regular; therefore, the Israel junction conditions
\cite{Israel:1966rt} must be fulfilled on the bubble wall. The
trajectory of the wall can be obtained easily considering the
coordinates in the 5-dimensional Minkowski space, where a de
Sitter spacetime can be visualized as the hyperboloid \cite{HyE}
\begin{equation}\label{hiperboloide}
-v^2+w^2+(x^i)^2=\alpha^2,
\end{equation}
with $\alpha=\sqrt{3/\Lambda}$ and $\Lambda$ is the  cosmological
constant. Due to the high level of symmetry of this space, one can
introduce different charts of coordinates in the hyperboloid
expressing the metric by different $3+1$ decompositions. These are
given by
\begin{equation}\label{metrica}
{\rm d}s^2=-{\rm d}t^2+a(t)^2\left[{\rm
d}\eta^2+f\left(\eta\right)^2 \,{\rm d}\Omega_{(2)}^2\right],
\end{equation}
where ${\rm d}\Omega_{(2)}^2={\rm d}\theta^2+\sin^2\theta {\rm
d}\phi^2$, with $0\leq\theta\leq\pi,\,0\leq\phi\leq 2\pi$. The
closed slicing, given by $a(t)=\alpha\,{\rm
cosh}\left(t/\alpha\right)$ and $f\left(\eta\right)=\sin\eta$,
with $-\infty<t<\infty$ and $0\leq\eta\leq\pi$, covers the whole
hyperboloid and leads to the Carter-Penrose (C-P) diagram through
$T=2\,{\rm arctan}\left[{\rm
exp}\left(t/\alpha\right)\right]-\pi/2$. The functions
$a(\hat{t})=\exp\left(\hat{t}/\alpha\right)$ and
$f\left(\hat{\eta}\right)=\hat{\eta}$ correspond to the flat
slicing with $-\infty<\hat{t}<\infty$ and $0\leq\hat{\eta}$. The
open slicing is given by $a(\tau)=\alpha\,{\rm
sinh}\left(\tau/\alpha\right)$, $f\left(\chi\right)={\rm
sinh}\chi$, $\tau\geq 0$ and $\chi<\infty$. We consider, without
lost of generality, that the bubble nucleates at $t=0$ and that it
is centered at $\eta=0$, where the coordinates of the closed
slicing have been taken. Therefore, following a similar procedure
to that considered in Ref.~\cite{Garriga:1997ef} for the flat
slicing, the trajectory of the wall in the outside space can be
given by fixing $w^*=D$. Therefore, this trajectory can be
expressed as
\begin{equation}\label{trayectoria}
\eta^*(t)={\rm arccos}\left[\frac{D}{\alpha
\cosh(t/\alpha)}\right],
\end{equation}
where $^*$ means evaluation on the wall\footnote{This trajectory
is the same as that obtained in Ref.~\cite{Lee:1987qc}, where a
different approach is followed.}. As it can be seen from condition
(\ref{hiperboloide}) and the relation between the coordinates of
the hyperboloid and those of the closed slicing \cite{HyE}, we
have $D<\alpha$ for $D>0$, implying that the trajectory is
well-defined. Therefore, the outside region can be described by
the closed slicing with $\eta^*\leq\eta\leq\pi$. The bubble wall
tends to $\pi/2$ when $t\rightarrow\infty$, see
Fig.~\ref{diagtray}. A bubble which nucleates with the minimal
possible size, $D/\alpha\rightarrow1$, will expand with maximal
velocity; although it must be noticed that the case $D=\alpha$
cannot be properly studied by using the thin-wall approximation.
Considering decay models leading to smaller values of this
quotient, one obtains bubbles with bigger initial sizes, tending
the trajectories of their walls to that of the smallest bubble for
large values of $T$.

The first junction condition, ${\rm d}s^2|_{{\rm wall}}={\rm
d}s_b^2|_{{\rm wall}}$ (where the subscript $b$ denotes the
quantities related to the inside space), can also be imposed in
the 5-dimensional space more easily than in the 4-dimensional one.
It leads to a trajectory of the wall in the inside space which
takes a similar form to that of the outside space,
Eq.~(\ref{trayectoria}), with
\begin{equation}\label{Db}
D_b^2=D^2+\alpha_b^2-\alpha^2,
\end{equation}
and to the following equation
\begin{equation}\label{1jc}
\sinh(t_b/\alpha_b)=\alpha/\alpha_b \sinh(t/\alpha),
\end{equation}
which must be fulfilled on the wall. Therefore, as seen from the
inside space, the bubble contains a region bigger than the future
light cone of its center (see Figs.~\ref{diagtray}). It must be
worth noticed that this light cone contains an infinite universe
inside when the open slicing is considered
\cite{Gott:1982zf,Bucher:1994gb}.

The second junction condition relates the value of the difference
of the extrinsic curvatures of both regions on the wall to the
surface density of the energy-momentum tensor on the wall
\cite{Berezin}. Therefore, the particular value of $D$ depends on
the value of both cosmological constants and on the particular
form of the potential which presents the two minima
\cite{Parke:1982pm}.
The fulfillment of this junction condition has led to some controversy about
the possible decay of a true
vacuum into a false vacuum \cite{Lee:1987qc}, since a false vacuum bubble able to grow
indefinitely must necessarily have emerged from an initial
singularity if the null energy condition is satisfied
\cite{Farhi:1986ty}.
Nevertheless, the consideration of some
scenarios \cite{Ansoldi:2007fw,Lee:2007dh} would lead to the
possible existence of unbounded false vacuum bubbles. Therefore,
we will impose no restriction on the relation between both
energies, in order to maintain our analysis as general as
possible.


On the other hand, there are a number of known properties of CMC
hypersurfaces in cosmological spacetimes satisfying the strong
energy condition \cite{Marsden}. Nevertheless, if the strong
energy condition is violated, as it occurs in a
de Sitter space with or without CDL bubble, then even less results are
available
(see Ref.~\cite{Beig:2005ef} and references therein for
information about developments in this particular field, and Ref.~\cite{Malec:2009hg} for recent studies considering different
spacetimes).
Nevertheless, we can notice that the hypersurfaces with CMC
in a de Sitter space with a thin-wall CDL bubble of de Sitter
space inside would also have CMC in a simple de Sitter space. The
closed slicing of a de Sitter space provides us with a simple CMC
foliation which covers the whole space, that is the constant
cosmic time foliation. But the foliation by
hypersurfaces with constant $t$ in the outside region and with
constant $t_b$ inside the bubble is not a CMC foliation.
Therefore, in order to find a CMC foliation of the considered
space, in the first place, we should solve the differential
equation implied by $K={\rm constant}$. However, it can be seen
that this is a non-trivial differential equation.

Let us reflect on the symmetry of the problem to find particular
solutions to the differential equation. A de Sitter hyperboloid is
a CMC 4-hypersurface in the 5-di\-men\-sion\-al Minkowski space.
Thus, one could think that the intersection of two 4-hypersurfaces
with CMC in a 5-di\-men\-sional Minkowski space produces a
3-hypersurface which would also have CMC, when considering this
hypersurface defined in the spacetime given by one of the original
4-hypersurfaces. Thus, taking this argument and the symmetry of
the problem into account, one can consider the
simplest\footnote{The spacelike hyperboloids also have
$K_{(4)}={\rm constant}$.} 4-dimensional hypersurfaces with CMC in
a Minkowski space; those are the hyperplanes which can be seen as
lines in a $(v,\,w)$-section of the space, i. e.
$\Sigma_4:\,\,v=b\,w+a,$ where $b$ and $a$ are arbitrary
constants, which describe the slope and the $v$-intercept,
respectively. It must be worth noticed that, in the 5-dimensional
Minkowski space, the hyperplanes are related by boots with those
4-hypersurfaces which lead to the constant cosmic time foliation
of the closed slicing when they intersect the hyperboloid;
therefore, they are, of course, CMC in the 5-dimensional space
(with $K_{(4)}=0$). Nevertheless, this is not necessarily ensuring
that their intersections with the hyperboloid lead to CMC
3-hypersurfaces in the 4-di\-men\-sion\-al de Sitter space; that
is precisely what we are arguing. Therefore, we must calculate the
trace of the extrinsic curvature of the 3-hypersurfaces,
$K=K_{(3)}\neq K_{(4)}$, in order to check
whether they really have CMC in a de Sitter space. The
intersections of the hyperplanes and the hyperboloid lead to the
hypersurfaces
\begin{equation}\label{cosh}
\cosh(t/\alpha)=\frac{ba\cos\eta+\sqrt{a^2+1-b^2\cos^2\eta}}{1-b^2\cos^2\eta},
\end{equation}
in a de Sitter space; where we have ruled out the solution with a
minus sign multiplying the square root, because they must simplify
to the correct constant value for $b=0$ (which is the case of the
constant $t$ foliation). The trace of the extrinsic curvature of
these 3-hypersurfaces is
\begin{equation}\label{Kout}
K=-3\frac{a}{\alpha\sqrt{a^2+1-b^2}},
\end{equation}
which is constant for each hypersurface. Thus, these are
spherically symmetric CMC hypersurfaces in a de Sitter space.

In the second place, as the previous procedure is valid for any de
Sitter space, we consider that the hypersurfaces are given in the
inside space by a similar expression. Therefore, the hypersurfaces
are CMC throughout the whole space only if $K=K_b$, that is
\begin{equation}\label{c1}
\frac{a}{\alpha\sqrt{a^2+1-b^2}}=\frac{a_b}{\alpha_b\sqrt{a_b^2+1-b_b^2}}.
\end{equation}

In the third place, the hypersurfaces must be regular. Therefore,
we have to impose: (i) the hypersurfaces must fulfill the first
junction condition on the bubble wall, and (ii) the scalar product
of the orthonormal vector to the CMC hypersurface and the
orthonormal vector to the wall at the intersection of both
hypersurfaces must be constant\footnote{This condition follows
from the requirement of the existence of a regular orthonormal
vector to the hypersurfaces. As we cannot compare vectors defined
in different spaces, we take the scalar product of the orthonormal
vector to the hypersurface and the orthonormal vector to the
bubble wall in the inside and outside regions. This condition is,
of course, less restrictive, but it allows us to compare
quantities of different spaces.}. These conditions imply:
\begin{equation}\label{c2}
bD+a\alpha=b_bD_b+a_b\alpha_b,
\end{equation}
and
\begin{eqnarray}\label{c3}
\frac{a
\frac{D}{\alpha}+b}{\sqrt{(1-\frac{D^2}{\alpha^2})(a^2+1-b^2)}}=\frac{a_b
\frac{D_b}{\alpha_b}+b_b}{\sqrt{(1-\frac{D_b^2}{\alpha_b^2})(a_b^2+1-b_b^2)}}.
\end{eqnarray}

Thus, we have a system of three equations, Eqs.~(\ref{c1}),
(\ref{c2}) and (\ref{c3}), with four unknown quantities, $\{a,\,
b,\, a_b,\, b_b\}$. This system has two sets of
solutions\footnote{There are two more sets of solutions, which
depend on the quantity $\sqrt{D^2-\alpha^2}$. Nevertheless, those
solutions have no physical meaning, since some parameters
take complex values.}, which are
\begin{eqnarray}\label{sol1}
a=-b\frac{\alpha}{D},\,\, a_b=-b\frac{\alpha_b}{D},\,\,
b_b=b\frac{D_b}{D},
\end{eqnarray}
and
\begin{eqnarray}\label{sol2}
&&a=-b\frac{\alpha(\alpha^2-\alpha_b^2)}{-2\alpha\alpha_bD_b+D(\alpha^2+\alpha_b^2)},{}
\nonumber\\
&&{}
a_b=b\frac{\alpha_b(\alpha^2-\alpha_b^2)}{-2\alpha\alpha_bD_b+D(\alpha^2+\alpha_b^2)},{}
\nonumber\\
&&{}b_b=b\frac{D_b(\alpha^2+\alpha_b^2)-2D\alpha\alpha_b}{-2\alpha\alpha_bD_b+D(\alpha^2+\alpha_b^2)}.
\end{eqnarray}
Each set of solutions describes a different foliation, being each
hypersurface given by a particular value of $b$. Foliation I,
Eq.~(\ref{sol1}), can be studied by using general considerations.
Whereas it is not possible to perform a detailed study about
foliation II, Eq.~(\ref{sol2}), without restricting to a
particular decay model. Nevertheless, it can been noticed, by
studying both foliations for the same particular values of the
parameters, that foliation II covers a region of the space smaller
than that covered by foliation I, at least for those values.
Therefore, foliation I and II are not equivalent. It must be
emphasized that these foliations are only the foliations coming
from a particular kind of CMC hypersurfaces in the 5-dimensional
space, the 4-hyperplanes. If the argument which we have included
would be valid in general, then it could help us to find more CMC
foliations. Moreover, it should be kept in mind that it is also
possible, at least in principle, that there would be foliations by
CMC hypersurfaces of a de Sitter space which are not produced by a
family of CMC 4-hypersurfaces in the 5-dimensional Minkowski space
intersecting the hyperboloid.

\begin{figure}
\begin{center}
\includegraphics[width=0.8\columnwidth]{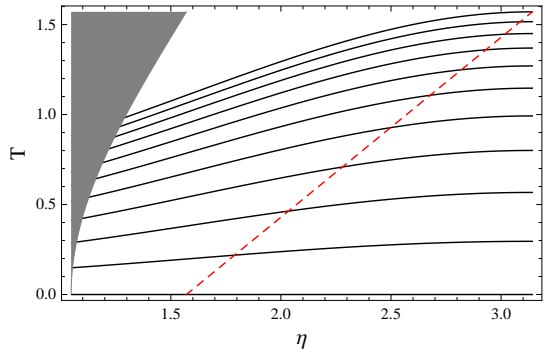}
\end{center}
\caption{CMC hypersurfaces (continuous lines) in the region
$T\geq0$ of the C-P diagram of the outside space for a model with
$\lambda=0.5$. The gray region is not part of this space. The
closed slicing covers the whole diagram, whereas the flat slicing
only covers the region of this diagram on the left of the dashed
line. In this figure the geodesics of the congruence would be
vertical straight lines covering the whole diagram.} \label{dos}
\end{figure}

We can express the hypersurfaces of foliation I in the outside
space, taking into account Eqs.~(\ref{cosh}) and (\ref{sol1}), as
\begin{eqnarray}\label{ch1}
\cosh\left(t/\alpha\right)=\frac{-b^2\cos\eta+\sqrt{b^2+\lambda^2(1-b^2\cos^2\eta)}}{\lambda(1-b^2\cos^2\eta)},
\end{eqnarray}
where $\lambda\equiv D/\alpha$, with $0<\lambda<1$. The
hypersurfaces are well-defined for $b^2\leq1$. As we are only
interested in the no-contracting region, we consider $b\leq0$
(which corresponds to $K\leq 0$); therefore $-1\leq b\leq0$. It
can be noticed that the hypersurface given by $b=-1$ diverges at
$\eta=\pi$, and every hypersurface intersects the wall at a finite
$t$. Thus, the foliation covers an infinite region of the outside
space, but it also avoids an infinite part. The foliation covers a
``larger'' region of the space for smaller values of $\lambda$,
that is for bubbles nucleated with bigger initial sizes. In
Fig.~\ref{dos} we show the CMC hypersurfaces in the C-P diagram of
the outside space for a particular decay model.

\begin{figure}
\begin{center}
\includegraphics[width=0.8\columnwidth]{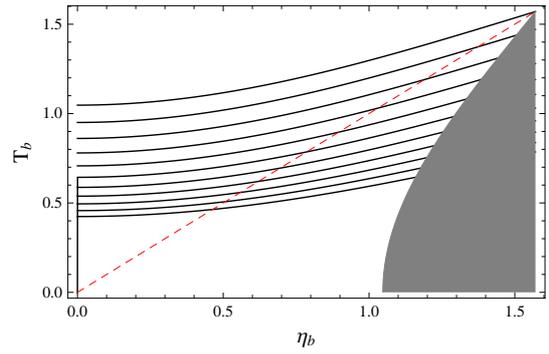}
\end{center}
\caption{C-P diagram of the inside space
for different models with $\lambda_b=0.5$ and $\beta$ in the
interval $[0,1]$. The gray region is not part of this space. The
maximal CMC hypersurfaces (continuous lines) of those models,
which corresponds to $b=-1$, are shown. The upper maximal CMC
curve corresponds to $\beta=0$, and the consideration of bigger
values of $\beta$ would lead to the appearance of more lines on
the bottom. The closed slicing covers the whole diagram, whereas
the open slicing only covers the future light cone of the bubble
center (region of this figure on the left of the dashed line).
The geodesics of the congruence would be vertical straight lines
covering the region of the diagram given by $0\leq\eta_b\leq{\rm
arccos}\lambda_b$.} \label{CPinside}
\end{figure}

Those hypersurfaces can be described in the inside region by
\begin{eqnarray}\label{ch1b}
&&\cosh\left(t_b/\alpha_b\right)={}
\nonumber\\ {}&&
\frac{-b^2\lambda_b\cos\eta_b+\beta\sqrt{\beta^2+b^2(1-\lambda_b^2\cos^2\eta_b)}}{\beta^2-\lambda_b^2
b^2\cos^2\eta_b},
\end{eqnarray}
with $\lambda_b\equiv D_b/\alpha_b$, $0<\lambda_b<1$, and
$\beta\equiv D/\alpha_b$. Taking into account Eq.~(\ref{Db}), it
can be obtained that $0<\beta<\lambda<\lambda_b<1$, in the case of
a true vacuum bubble (TVB) which nucleates in a false vacuum
($\alpha_b>\alpha$); whereas if one considers a false vacuum
bubble (FVB, $\alpha_b<\alpha$), then
$0<\lambda_b<\lambda<\beta<\alpha/\alpha_b$. These hypersurfaces,
with $-1\leq b\leq0$, are well-defined in the considered range.
The foliation only covers a finite region inside the bubble. This
covered finite region is larger for smaller values of $\beta$. For
values of $\lambda_b\rightarrow1$, the foliation only covers a
small region of the inside space; this fact can be understood
thinking that in this case the initial size of the bubble is so
small that there is almost nothing to cover at $t_b=0$, and the
bubble grows so quickly that the foliation cannot come into the
inside space. For values of $\lambda_b\rightarrow0$, the foliation
covers a larger region of the inside space, because this space
almost corresponds to the whole causal connected region of the
bubble center (see Fig.~\ref{diagtray}). In Fig.~\ref{CPinside} we
show the behavior of the CMC hypersurfaces at the end of the
foliation ($b=-1$) in the C-P diagram of the inside space for
different decay models.

It must be worth noticed that this foliation covers the range
$K\in [-3/\alpha,\,0]$, which is the same interval covered by the
constant cosmic time foliation for the no-contracting region of a
de Sitter space. Therefore, if some quantity proportional to $K$
can be interpreted as a preferred time, the York time \cite{York},
then this foliation covers the same interval as that covered by
the constant $t$ foliation in a de Sitter space, at least in
principle.


In order to understand whether the foliation could cover the
existence of most observers in our part of spacetime, we study how
far a particular congruence of geodesics can go in cosmic time
before reaching the end of the foliation. We can consider
the congruence of geodesics orthogonal to the hypersurface
$S_0:\,t=0$, with the same boundaries as the outside region at
$t=0$. Thus, the congruence, $\gamma_s^\mu(t)$, has affine
parameter $t$, which is just the cosmic time of the closed
slicing. If we suppose that $\Omega$ is slightly bigger than 1,
then the geodesics advance until a cosmic time which can be
obtained by calculating the intersection of the geodesics and the
maximal hypersurface\footnote{In this paper what we call maximal
hypersurface must not be confused with other meanings, as that used
in Ref.~\cite{York}.}, corresponding to $b=-1$; this is
\begin{equation}\label{intgeo}
t_{{\rm max}}=\alpha \, {\rm arccosh}
\left(\frac{-\cos\left(s\right)+\sqrt{1+\lambda^2\sin^2\left(s\right)}}{\lambda\,\sin^2\left(s\right)}\right).
\end{equation}
Nevertheless, if $\Omega$ is just $1$, as suggested by the
observational data \cite{de Bernardis:2000gy}, then we must obtain
$\hat{t}\left(t_{{\rm max}},s\right)$ and the particular cosmic
time at which each geodesic comes into the flat slicing, since
that slicing is not geodesically complete (see Fig.~\ref{dos}). It
is interesting to consider some particular geodesics in detail. In
the first place, we take the geodesic at the upper boundary of the
outside region, i. e. $s=\pi$. We can see, through
Eq.~(\ref{intgeo}), that $t_{{\rm max}}\rightarrow\infty$ when
$\eta\rightarrow\pi$, independently of the value of $\lambda$;
therefore, the existence of an observer with ideal infinite life
and cosmic time $t$ whose trajectory is defined by this geodesic
is completely covered by the foliation. On the other hand, this
geodesic only covers one point of the flat slicing, being covered
itself by the foliation at this point, which is the spacelike
infinity $i^0$. In the second place, we consider the complete
geodesic with the smallest value of $s$, that is $s=\pi/2$. This
geodesic advances in the closed slicing until $t_{{\rm
max}}=\alpha\,{\rm arccosh}\left(\sqrt{\lambda^{-2}+1}\right)$,
which is an enormously big time for small values of $\lambda$ and
tends to $t_{{\rm max}}=0.8812\alpha=1.5265\Lambda^{-1/2}$ for
$\lambda\rightarrow 1$. In the flat slicing it advances
from a point in the past timelike infinity, $I^-$, until a point
given by $\hat{r}=\alpha\sqrt{1+\lambda^2}$ and
$\hat{t}=-\alpha\log\lambda$, which leads to an infinitely big
(small but non-vanishing) cosmic time for small (large) values of
$\lambda$. Therefore, if we are in a universe with a value of $\Omega$ slightly larger than
$1$ or just $1$, then our existence is generally covered by
foliation I. Thus, the use of this CMC foliation will consistently lead to a non-null probability for the events that
we measure, which could be arbitrarily large by choosing a suitable measure, allowing us to be
typical observers whose observations are likely.


We are considering a scenario which is compatible with our accelerated Universe, if we are placed in the outside de Sitter space.
Nevertheless, let us consider for a moment that we are placed in the space inside the bubble. In this case, our Universe would nucleate
in another universe, being $\tau_b$ our cosmic time \cite{Gott:1982zf}.
Therefore, we can also consider a similar congruence of geodesics in the inside space, $\gamma_p^\mu(t_b)$.
Each geodesic advances until an affine parameter expressed through
\begin{eqnarray}
&&\cosh\left(t_{b\,{\rm max}}/\alpha_b\right)={}
\nonumber\\ {}&&
\frac{-\lambda_b\cos\left(p\right)+\beta\sqrt{\beta^2+1-\lambda_b^2\cos^2\left(p\right)}}{\beta^2-\lambda_b^2\cos^2\left(p\right)}.
\end{eqnarray}
In this framework, one must consider
the geodesic with $p=0$, which advances until an affine parameter
given by
\begin{equation}
t_{b\,{\rm max}}=\alpha_b\,{\rm
arccosh}\left(\frac{-\lambda_b+\beta\sqrt{\beta^2+1-\lambda_b^2}}{\beta^2-\lambda_b^2}\right),
\end{equation}
being covered by the foliation. This geodesic is completely
contained in the open slicing and describes the temporal evolution
of $\chi=0$. It advances from $\tau_b=0$ until $\tau_{b\,{\rm
max}}=t_{b\,{\rm max}}$, being $\tau_{b\,{\rm max}}$ enormously
big (although bounded) for small values of $\lambda_b$, and arbitrarily
small for $\lambda_b\rightarrow1$. Moreover, considering a fixed
value of $\lambda_b$, $\tau_{b\,{\rm max}}$ is bigger for smaller
values of $\beta$; thus, the foliation covers a larger part of our
existence for TVB models. On the other hand, other geodesics could
come into the open slicing being still covered by the foliation.
This is the case of the geodesic at the other boundary of $S_0^b$,
$p={\rm arccos}\lambda_b$, if $t_{b\,{\rm max}}>\alpha_b\,{\rm
arccosh}\left(\lambda_b^{-1}\right)$, with
\begin{equation}
t_{b\,{\rm max}}=\alpha_b\,{\rm arccosh}\left(\frac{-\lambda_b^2+\beta\sqrt{\beta^2+1-\lambda_b^4}}{\beta^2-\lambda_b^4}\right).
\end{equation}
This condition implies
$\beta<\lambda_b\left(1-\lambda_b\right)/\sqrt{1-\lambda_b^2}$;
thus, it can be fulfilled only by some TVB models. In this case,
the geodesic advances in the open space from $\tau_b=0$ until
\begin{equation}
 \tau_{b\,{\rm max}}=\alpha_b\,{\rm arccosh}\left(\lambda_b\frac{-\lambda_b^2+\beta\sqrt{\beta^2+1-\lambda_b^4}}{\beta^2-\lambda_b^4}\right),
\end{equation}
which takes finite and non-vanishing values. $\tau_{b\,{\rm max}}$
is bigger for smaller values of $\lambda_b$, tending to vanish for
$\lambda_b\rightarrow1$. Therefore, if the considered spacetime would be taken to describe the 
birth of our Universe, then the foliation would only cover our existence form some particular
decay models. Thus, in this hypothetical case, the use of this CMC foliation to count the events in the spacetime 
would lead to an undesirable result in some decay models, that is, our observations would not be likely.


\begin{figure}
\begin{center}
\includegraphics[width=0.8\columnwidth]{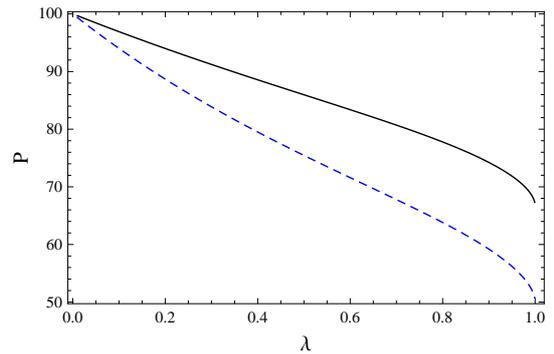}
\end{center}
\caption{These curves show the behavior of ${\rm
P}\left(\lambda\right)$ considering the whole diagram (continuous
line) and only the region defined by the flat slicing (dashed
line). In both cases ${\rm P}\left(\lambda\right)$ decreases for
increasing values of $\lambda$. Nevertheless, its minimum is above
$50$, which means that even in the worst case the foliation would
cover most of the outside space.} \label{PP}
\end{figure}

\begin{figure}
\begin{center}
\includegraphics[width=0.8\columnwidth]{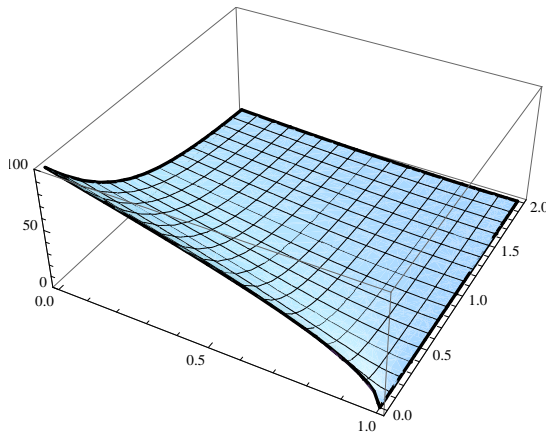}
\end{center}
\caption{Portion of the future light cone of the bubble center
covered by the CMC foliation in the inside space. We have only
shown values of $\beta$ in the interval $[0,\, 2]$.}
\label{Portcono}
\end{figure}

Up to now we have pointed out that the use of foliation I could lead to
satisfactory results regarding our typicality. However, we must also consider
whether we are taking a suitably approximation when considering this foliation.
Therefore, we should study the portion of the
C-P diagram which is covered by the foliation. In the first place,
we consider the whole outside space. We can roughly estimate the
area being covered by the foliation in the diagram by
approximating the area covered by the curves by the area covered
by the lines which have the same endpoints. In the second place,
we consider only the region of the diagram defined by the flat
slicing, assuming $\Omega=1$. In Fig.~\ref{PP} we show the
behavior of ${\rm P}\left(\lambda\right)=100\times{\rm A}_{\rm
cov}/{\rm A}_{\rm tot}$ in both cases. The foliation covers most
of the outside space (${\rm P}\left(\lambda\right)>50\%$)
independently of the particular decay model, considering both
$\Omega>1$ and $\Omega=1$.
On the other hand, following a similar procedure as that applied in
the outside space, one can obtain the portion of
the future light cone of the bubble center covered by the foliation, ${\rm
P}_b\left(\lambda_b,\,\beta\right)$, see
Fig.~\ref{Portcono}. It can be noticed that the foliation is able to cover
most part of the inside space only for very particular TVB decay
models. However, that portion takes larger values considering the whole inside region, and the total portion, 
considering both regions, takes values even larger. Therefore, in the cases that this total portion is larger than $50\%$,
one can suitably approximate the events of this spacetime by using this CMC foliation, and, at the end of the day, one could
even conclude that we are typical observers by taking a suitable measure when defining probabilities. In the other cases,
it should be worth noticed that to define probabilities for the occurrence of the events in this spacetime,
one must not only add up the events of each hypersurface taking a certain measure, but also count the contribution
of each region of the whole space; therefore, the use of this foliation could
also provide us with an accurate approximation in those cases, if our region contributes more to the path integral,
as it has been already suggested \cite{Page:2008zh}.

\begin{acknowledgement}
The author thanks the University of Alberta for hospitality and is indebted to Don N.~Page for suggesting me
this subject and for comments crucial for the development of this
work. I also thank Pedro F.~Gonz\'alez-D\'{\i}az and Salvador
Robles-P\'erez for useful discussions and gratefully acknowledge
the financial support provided by the I3P framework of CSIC and
the European Social Fund. This work was supported by MICINN under
research project no. FIS2008-06332.
\end{acknowledgement}


\end{document}